# Geometry and quantum delocalization of interstitial oxygen in silicon

Emilio Artacho, Arturo Lizón-Nordström, and Félix Ynduráin
*Instituto de Ciencia de Materiales Nicolás Cabrera and Departamento de Física de la Materia Condensada, C-III*
*Universidad Autónoma de Madrid, 28049 Madrid, Spain.*

The problem of the geometry of interstitial oxygen in silicon is settled by proper consideration of the quantum delocalization of the oxygen atom around the bond-center position. The calculated infrared absorption spectrum accounts for the 517 and 1136 cm$^{-1}$ bands in their position, character, and isotope shifts. The asymmetric lineshape of the 517 cm$^{-1}$ peak is also well reproduced. A new, non-infrared-active, symmetric-stretching mode is found at 596 cm$^{-1}$. First-principles calculations are presented supporting the nontrivial quantum delocalization of the oxygen atom.

Although the atoms in a solid are delocalized, one can talk about atomic positions whenever the harmonic approximation applies, meaning that the atoms are trivially delocalized around potential minima according to the quantization of normal modes. This picture does not hold when the potential is not harmonic for the lowest quantum states. The maximum probability of finding an atom may be away from a potential minimum.[1] A more dramatic deviation from the usual picture arises when one atom is delocalized among several equivalent minima related by symmetry. This has been observed with very light atoms, like H or Li.[2] This nontrivial delocalization prevents the classical assignment of a position to the atom.

Interstitial oxygen in crystalline silicon ($O_i$) is an example of this behavior. This very common and important impurity has been extensively studied for more than thirty years.[3] It was soon described in terms of the oxygen atom breaking a bond between two silicon atoms, forming a Si–O–Si bridge. In a seminal work by Hayes and Bosomworth[4] the nontrivial delocalization of the oxygen atom was pointed out as a result of a far-infrared-absorption investigation of the problem. This delocalization is in the plane perpendicular to the original Si–Si bond line, around the bond-center (BC) position. The potential seen by oxygen is axially symmetric around that line, corresponding to the $D_{3d}$ point group. The far infrared (FIR) experiments demonstrate that this symmetry is essentially not broken, the oxygen being nontrivially delocalized around the axis. To a good approximation, the potential seems to be invariant under any rotation around this axis.

In addition to that angular delocalization, the analysis in ref. 4 of the FIR spectra suggests that the potential is highly anharmonic in the radial direction, too. Their proposed model potential displays a local maximum at the BC site and a local minimum 0.22 Å off the axis, defining an energy barrier of approximately 30 cm$^{-1}$. The lowest FIR peak measured is at 29.3 cm$^{-1}$, the final state for this transition lying well above the maximum of the potential. This indicates a nontrivial quantum delocalization of the oxygen *also* in the radial motion. Furthermore, when considering the anharmonic coupling of these low energy anharmonic excitations (LEAE) with the asymmetric stretching mode of the defect, Yamada-Kaneta *et al.*[5] have shown that the energy barrier reduces down to be considerably lower than the energy of the ground state. It can be shown that such a situation leads to a ground state for the oxygen motion with *maximum probability density at the BC site in spite of being at a maximum of the potential,* as long as the ground-state energy is higher than the barrier. The position of the minimum of the potential looses its relevance to the definition of the geometry of the system.

However, the simple geometry picture related to the potential minimum has been maintained in most of the subsequent work, both experimental[6–8] and theoretical,[9–13] where the discussions rely on a 'puckered geometry', *i.e.*, a Si–O–Si angle smaller than 180°. This misleading geometry picture has prevented the achievement of a global understanding of the problem in spite of the fact that most of the relevant pieces of information were already present in the literature. In this Letter we present a comprehensive picture of the $O_i$ system. The most striking consequence of our analysis is the finding of two vibration modes, with different character and symmetry, in a frequency region (500 - 600 cm$^{-1}$) where only one mode has been discussed so far.

In addition to the LEAE and other combination modes related to them,[4,7] two main features have been experimentally observed in the infrared absorption spectra of $O_i$,[3,7] namely at 517 and 1136 cm$^{-1}$. Although the character of the modes involved in the 1136 cm$^{-1}$ band is generally accepted, there is still some controversy concerning the characterization of the 517 cm$^{-1}$ mode.[5,6] We have performed a model calculation of the vibration modes and infrared absorption of the system. The calculation assumes a classical Born potential between nearest-neighbor atoms of the form[14]

$$V_{ij} = \frac{(\alpha - \beta)}{2}([\mathbf{u}(i) - \mathbf{u}(j)] \cdot \mathbf{r}_{ij})^2 + \frac{\beta}{2}[\mathbf{u}(i) - \mathbf{u}(j)]^2$$

where $\mathbf{u}(i)$ stands for the displacement of atom labeled $i$ and $\mathbf{r}_{ij}$ represents the unit vector joining atoms labeled $i$ and $j$. The central and non-central forces are given by $\alpha$ and $\beta$, respectively. The forces between silicon atoms are those of bulk silicon[14] and for the Si–O bond we take values of SiO$_2$, namely $\alpha = 4.8 \cdot 10^5$ and $\beta = 0.3 \cdot 10^5$ dynes/cm. The quantization of the normal modes obtained from this potential does not correctly describe the LEAE nor the fine structure of the asymmetric stretching band found in the spectra. These two points are well treated in the literature[4,5] and therefore will not be considered here. Results will be presented only for modes with frequency above 400 cm$^{-1}$.

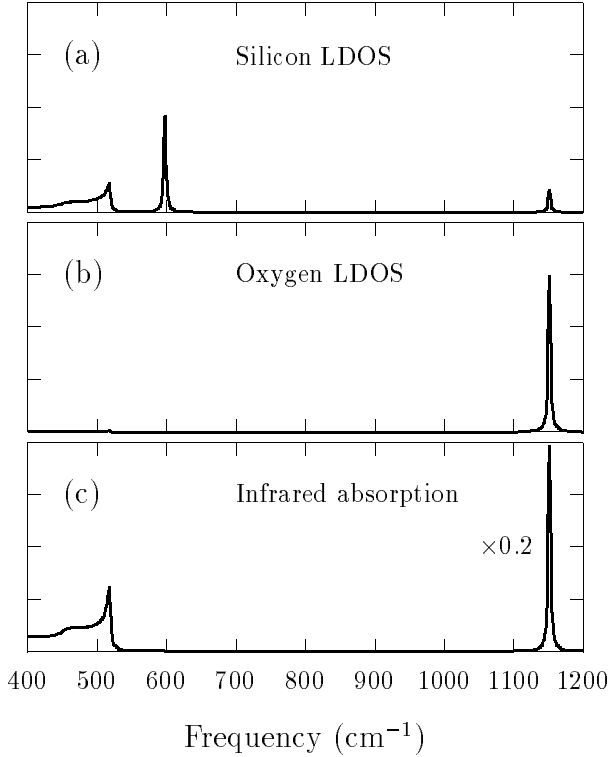

FIG. 1. Vibrations at the O$_i$ center. (a) Density of vibration modes projected on the Si atoms neighbors to O. (b) Density of modes projected on the O atom. (c) Infrared absorption. LDOS stands for local density of states.

The vibration modes as well as the correlation functions are calculated by means of the cluster-Bethe-lattice approximation.[14] This allows the calculation of the infrared absorption which is given by:[14]

$$\alpha(\omega) \propto -\omega \, \text{Im} \left[ \text{Tr} \left\{ \sum_{i,j} \mathbf{Q}_i \cdot \mathbf{G}_{ij}(\omega) \cdot \mathbf{Q}_j \right\} \right]$$

where $\mathbf{G}_{ij}$ stands for the Green's function tensor between atoms labeled $i$ and $j$ and $\mathbf{Q}_i$ is the charge tensor of atom $i$ whose elements are obtained from first-principles calculations (see below).

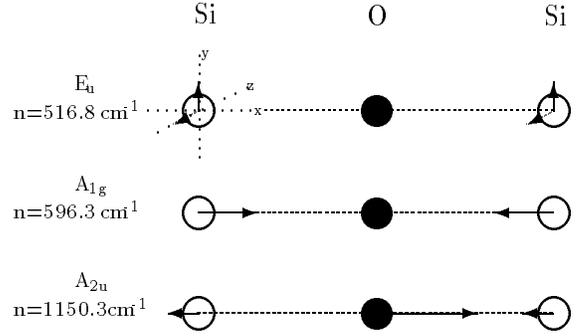

FIG. 2. Modes associated to the main features of the spectrum of Fig. 1. Relative amplitudes of atomic motion are shown as arrow lengths.

Since the maximum probability of finding the oxygen atom is at the BC position, the classical harmonic potential that better describes the system should have its oxygen equilibrium position at the BC site. This gives a correct representation of the $D_{3d}$ symmetry group (any classical puckered geometry would not). For the Si–O distance we take the value of 1.56 Å as indicated by total energy calculations (see below). This distance implies an important outwards relaxation of the silicon atoms nearest to oxygen. Results of the calculations for the oxygen and silicon local density of modes as well as for the infrared absorption spectrum are shown in Fig. 1. The modes associated to the main features of the vibration spectrum are indicated in Fig. 2. There are several points worth mentioning about these results:

i) The calculated infrared absorption agrees with the experimental results with one main peak at 1150 cm$^{-1}$ (in our calculation this a single peak since no interaction with LEAE is considered) and a much weaker peak at 517 cm$^{-1}$. The peak at 1150 cm$^{-1}$ corresponds to an asymmetric-stretching $A_{2u}$ vibration mode.

ii) The mode at 517 cm$^{-1}$ corresponds to a transversal motion of the silicon atoms nearest to oxygen with negligible participation of the oxygen atom. This $E_u$ mode is a defect-induced silicon mode originated by the fact that, due to the presence of oxygen, the silicon backbonds are in a more planar situation than in the tetrahedral configuration. This produces a resonance-like edge peak at the top of the silicon vibration spectrum resulting in an asymmetric line (see ref. 7) as it is seen in detail in Fig. 3. It mainly involves stretching of the Si backbonds. This line is, therefore, essentially independent of the Si–O interaction parameters. Our characterization of the mode agrees with Stavola's determination of the orientation of the transition moment for the band,[6] but disagrees with his interpretation concerning the Si–O stretching character of the mode.

iii) In the silicon local density of states in Fig. we observe a mode at 596 cm$^{-1}$ which has no infrared activity. This mode corresponds to a symmetric stretching $A_{1g}$ vibration of the silicon atoms nearest to oxygen and it is the symmetric counterpart of the mode at 1150 cm$^{-1}$.

Its infrared activity is strictly zero because of symmetry.

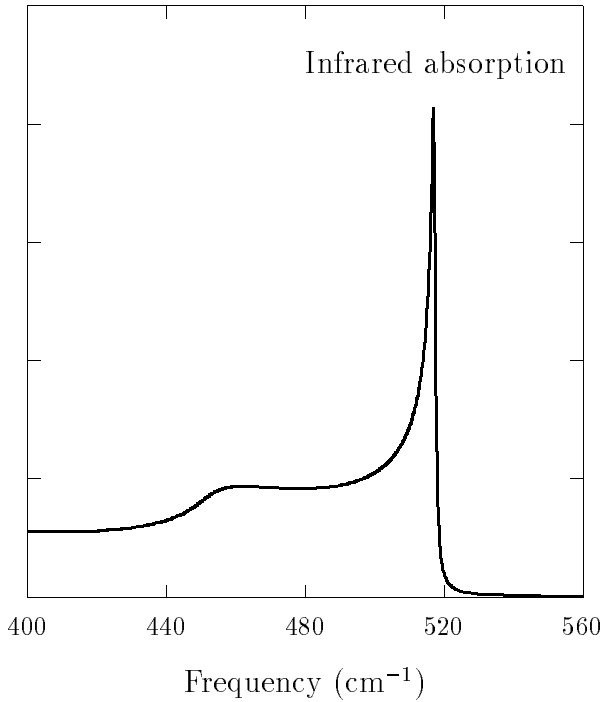

FIG. 3. Detail of the 517 cm$^{-1}$ resonance in the infrared absorption spectrum.

iv) The only way that the mode at 596 cm$^{-1}$ can be manifested experimentally in infrared measurements under the symmetry conditions is by combination with other infrared-active modes. In particular its combination with the 1150 cm$^{-1}$ mode is responsible for the experimentally observed band at 1700 cm$^{-1}$. The existence of a mode at around 600 cm$^{-1}$ was already speculated by Yamada-Kaneta et al.[5] in their interpretation of this band.

An indirect way to show the existence of this mode is by looking at geometrical configurations where the $D_{3d}$ symmetry is broken. In particular, the experimentally observed peaks at 720 and 1000 cm$^{-1}$ induced by the presence of interstitial atomic oxygen at the Si(111) surface[15] correspond to the above discussed modes at 596 and 1150 cm$^{-1}$, respectively. Their frequency shift is consequence of a symmetry-broken configuration near the surface with a Si–O–Si bond angle much smaller than 180°.

v) For a more complete characterization of the above discussed modes we have calculated their isotope shifts. Results of the calculations along with the experimental results are displayed in Table I. The good overall agreement between theory and experiment stresses the correctness of our analysis.

To gain more insight into the actual form of the potential seen by oxygen we have performed a first-principles calculation of the total energy as a function of the atomic geometry. Previous calculations either disagree with the experimental situation[9–12] or do not present results for the potential.[13] We have considered the presence of an oxygen atom in a saturated Si$_8$ cluster. The total energy for different atomic configurations is obtained by means of full Hartree-Fock calculations. The technique has been described elsewhere.[12] In Fig. 4 we show the total energy as a function of the oxygen distance to the BC site. The upper curve in panel (a) is obtained by allowing only the silicon atoms nearest to oxygen to relax. They move outwards, 0.36 Å each, along the Si–Si line. The lower curve in the same panel has been obtained by allowing both first and second nearest-neighbor atoms to relax. Nearest-neighbor relaxation is now 0.41 Å, second nearest neighbors relaxing 0.06 Å, essentially in the same direction.

This result stresses the importance of the relaxation of second nearest-neighbor atoms, which accounts for the substantial lowering (eventual disappearing) of the energy barrier in the radial motion of the oxygen atom. Relaxations to further neighbors are quantitatively important for the potential (it shifts down in energy), but its anharmonic character around the BC site remains. This is a result of the competition between the resistance against bending of the Si-$sp^3$ hybrids and the very large compressive stress the defect is under,[16] which cannot be relaxed no matter how far the defect-induced strain propagates.

TABLE I. Isotope shifts of the modes shown in Fig. 2. Experimental data are taken from refs. 4 and 5 (for $A_{1g}$ data for the 1136 and the 1700 cms$^{-1}$ bands are substracted).

|  | $E_u$ (516.8 cm$^{-1}$) | | $A_{1g}$ (596.3 cm$^{-1}$) | | $A_{2u}$ (1150.3 cm$^{-1}$) | |
|---|---|---|---|---|---|---|
|  | Exp. | Th. | Exp. | Th. | Exp. | Th. |
| $^{28}$Si$^{16}$O$^{28}$Si | 0.0 | 0.0 | 0.0 | 0.0 | 0.0 | 0.0 |
| $^{28}$Si$^{16}$O$^{29}$Si | – | 0.2 | 4.6 | 5.0 | 1.8 | 2.3 |
| $^{28}$Si$^{16}$O$^{30}$Si | – | 0.5 | 9.7 | 9.7 | 3.6 | 4.5 |
| $^{28}$Si$^{17}$O$^{28}$Si | 0.0 | 0.0 | – | 0.0 | 27.0 | 26.0 |
| $^{28}$Si$^{18}$O$^{28}$Si | 0.0 | 0.0 | – | 0.0 | 51.6 | 49.7 |

The second curve in Fig. 4 (a) is essentially flat around the origin for the energy scale displayed in the figure. The unambiguous determination of the shape of this potential in the meV range is beyond the capabilities of any first-principles method. This is indicated by the shaded region in the figure. The maximum barrier at the symmetric position compatible with our computational scheme is shown in Fig. 4 (b). Fig. 4 also shows the relevant energy scale indicating with a bar the energy of the lowest experimental transition within the LEAE.

There are two important qualitative consequences that can be drawn from these results. First, for the scale of energies relevant to the physical properties considered here, the total-energy curve represents an anharmonic potential. Second, the maximum barrier obtained is smaller than the lowest oxygen energy level and, therefore, the ground-state wave-function of oxygen has a maximum at the central position, as was considered in the calculation of the vibration modes.

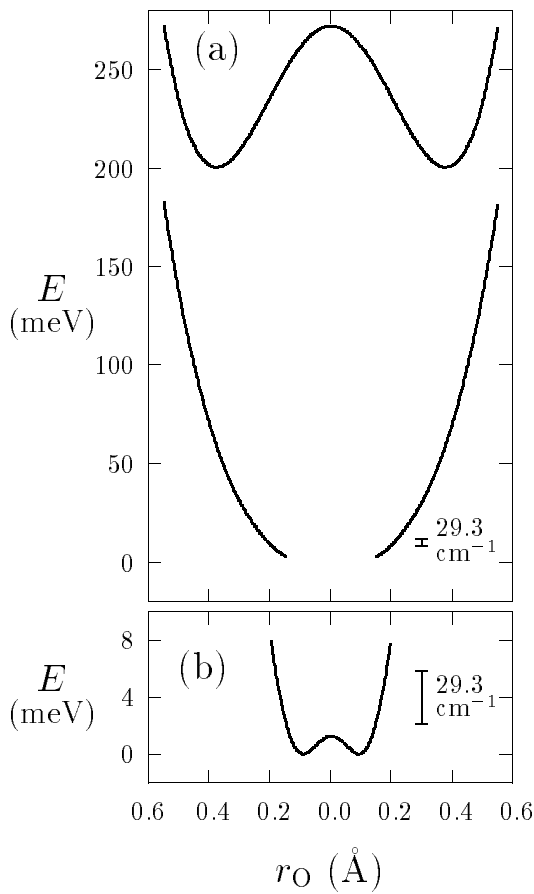

FIG. 4. *Ab initio* potential seen by the oxygen atom as a function of the distance to the BC site perpendicular to the symmetry axis. In (a), the upper (lower) curve is obtained with relaxations up to first (second) nearest neighbors of oxygen. The shaded area indicates a region of uncertainty in the calculations. Curve in (b) shows the maximum barrier we can achieve within our calculation scheme (1.3 meV). In both panels the bar shows the energy for the lowest experimentally observed transition.

In conclusion, in this work we have presented a consistent interpretation of the geometry and vibrations of interstitial oxygen in silicon. The quantum delocalization of oxygen around the BC site indicates the incorrectness of the broadly accepted representation of a puckered Si-O-Si bond with a defined geometry. A study including LEAE and the $A_{2u}$ and $A_{1g}$ modes interacting to obtain more realistic information about the potential from the experimental spectra is in progress and will be reported elsewhere.[16] Recent experiments on interstitial oxygen in germanium[17] are particularly interesting in the light of this contribution. They are being analyzed at the moment.[16]

We are indebted to M. Cardona and G. Gómez-Santos for stimulating discussions about this problem and to J. A. Torres. This work has been supported in part by the Dirección General de Investigación Científica y Tecnológica (DGICYT) through grant PB92-0169.